\documentstyle[12pt]{article}

\catcode`\@=11

 \@addtoreset{equation}{section}

\begin{document}

\baselineskip 24pt

\title{\large{\bf  EXACT RESULTS IN  SOFTLY BROKEN  \\
 SUPERSYMMETRIC MODELS}}

\author{
Nick Evans\thanks{nick@zen.physics.yale.edu} ,
Stephen D. H. Hsu\thanks{hsu@hsunext.physics.yale.edu}
\/ and Myckola Schwetz\thanks{ms@genesis2.physics.yale.edu} \\ \\
Department of Physics  \\   Yale University
\\ New Haven, CT 06520-8120   }

\date{}

\maketitle

\begin{picture}(0,0)(0,0)
\put(350,251){YCTP-P8-95}
\put(350,235){hep-th/9503186}
\end{picture}
\vspace{-24pt}

\begin{abstract}
We show how recent exact results in supersymmetric theories
can be extended to models which include {\it explicit} soft
supersymmetry breaking terms. We thus derive new exact results
for non-supersymmetric models.
\end{abstract}

\renewcommand\thefootnote{\fnsymbol{footnote}}

\section{Introduction}
The superpotential in the Wilsonian effective Lagrangian of a
supersymmetric
(SUSY) theory is holomorphic in the fields
and invariant under any gauged or global symmetries of the theory
\cite{SV,CERN}.
Recently Seiberg \cite{S1} has used these properties
to prove non-perturbative non-renormalization theorems in
SUSY theories and to explicitly explore the vacuum structure
of strongly interacting SUSY gauge theories. One of his
important tools
is to embed the theory under study in a larger SUSY
model in which the couplings of the original theory are treated as
chiral superfields (i.e. supersymmetric spurions).
The coefficients of the
kinetic terms for these chiral `coupling
superfields' are later taken to infinity, so that they
become constant, non-propagating, background fields.
However, the superpotential of this larger theory
must still be holomorphic in the coupling fields. By imposing
a non-zero vacuum expectation value (vev) for
the lowest, or scalar, component of the coupling fields the
embedded theory is recovered.

It is known that adding soft breaking
terms to SUSY theories does not destroy the
perturbative
non-renormalization theorems \cite{GG}.
That is, soft breaking does not induce any
power divergences and leads only to logarithmic
corrections to the superpotential.
In this paper
we show that these non-renormalization
theorems continue to be valid non-perturbatively as well.
Our method is, similarly, to embed the theory
we wish to study in a larger SUSY theory in which
again couplings are treated as chiral superfields.
However, now we couple each of the
coupling superfields to an
enlarged Wess-Zumino (WZ) sector which is an
O'Raifeartaigh model \cite{OR}. The superpotential is
again determined exactly by the symmetries of
the theory and holomorphy. The O'Raifeartaigh
sectors generate non-zero expectation values
for the $F$-terms of the coupling fields as
well as the scalar components. Then supersymmetry
is spontaneously broken and soft SUSY breaking
terms are generated in the embedded theory.
However, because the larger theory is exactly
supersymmetric, with SUSY only broken
spontaneously, the holomorphic properties of the
superpotential remain intact and lead to exact results.

We will thus derive exact results
 for the Wess-Zumino model and for SUSY gauge
theories in the presence of all possible soft SUSY breaking terms.
Among these soft SUSY breaking terms
are ones which may be used to give large masses to the
gauginos (e.g. gluinos and photino)
and the scalar components of matter superfields
(e.g. squarks and sleptons).
This will allow us to embed QED and
QCD in larger SUSY gauge models and then
to subsequently decouple all of the
superpartners without changing the holomorphic properties
of the potential.
We anticipate therefore that
our methods will be useful in investigating whether
Seiberg's results in SUSY QCD  (SQCD)
will be preserved in QCD. Work in this direction
is in progress.

\section{The Wess-Zumino Model}

\subsection{Non-Renormalization theory: one flavor WZ Model}

We begin by reviewing Seiberg's method in the Wess-Zumino model
\footnote{We use the conventions and notation
of Wess and Bagger \cite{WB} throughout this paper.}
\begin{equation}
\label{WZ}
L_0 = \Phi^{\dagger} \Phi \vert_D +
m \Phi^2 \vert_F + g \Phi^3 \vert_F
\end{equation}
where we have neglected the term with
a single power of $\Phi$ since it may be removed
by a shift in the field. The model may be embedded
in a larger theory where $m$ and $g$ are treated as fields:
\begin{equation}
\label{WZe}
L = L_0 +  \Lambda_m m^{\dagger} m \vert_D +
\Lambda_g g^{\dagger} g \vert_D
\end{equation}
where $\Lambda_m$ and $\Lambda_g$ are some scales.
The Wess-Zumino model is recovered
in the limit where the coupling fields $c = m,g$
are given vacuum expectation values (vevs)
in their scalar components $A_c$
and then $\Lambda_c \rightarrow \infty$. In this
limit the kinetic terms of the coupling fields  have
infinite coefficients. Hence any contribution
to the functional integral with field variation is suppressed,
and fluctuations of the coupling fields can be neglected.
Furthermore,
the vev of the scalar fields may be set to any value by appropriate choice
of a source term. Once the source is turned off, the
vev will remain fixed even if it does not minimize the potential
energy. This is easily seen by
examining the equation of motion for the scalar components
of the coupling fields, which have the form
\begin{equation}
\label{EOM}
\partial^2 A_c ~\sim~  \frac{1}{\Lambda_c}
{}~ \frac{\partial U}{\partial A_c},
\end{equation}
where $U$ is the scalar potential.

The enlarged model has a global
$U(1) \otimes U(1)_R$ symmetry under which the fields transform as
\begin{equation} \begin{array}{cccc}
 & U(1) & \otimes & U(1)_R\\
\Phi &  1 & & 1\\
m &  -2 & & 0\\
g &  -3 & & -1\end{array}
\end{equation}

Holomorphy and these U(1) symmetries
restrict the superpotential of the Wilsonian
effective Lagrangian to the form
\begin{equation}
\label{Weff}
W_{eff} ~=~ m\Phi^2 \hspace{.3cm} f\left( \frac{g \Phi}{m} \right)
\end{equation}
where $f $ is any holomorphic function.
The effective superpotential receives contributions from
perturbative 1PI loop
diagrams. It is easy to see that all such
diagrams are incompatible with the form of (\ref{Weff}).
For sufficiently small coupling $g$,
any remaining non-perturbative contributions to
$W_{eff}$ must approach zero faster than $g^N$ for
any $N$. We assume that they do so uniformly
for all $\Phi$ -- in other words the
non-perturbative corrections should be non-singular for all
field values at sufficiently weak coupling.
This assumption rules out any corrections, e.g., of the
form $e^{- 1/g}$ since to conform with (\ref{Weff}) they must
actually be of the form $e^{- m / g \Phi}$ and hence are singular
for $\Phi$ small and negative.
We conclude there are no corrections
compatible with the form of (\ref{Weff}) and
therefore the superpotential is not renormalized:

\begin{equation}
W_{eff} ~=~ m \Phi^2 + g \Phi^3 = W_{tree}  .
\end{equation}
Note that one could
also add to the Lagrangian  (\ref{WZe}) a term
\begin{equation}
\label{KK}
\delta L ~=~ ( K^{\dagger} K ~\Phi^{\dagger} \Phi )\vert_D
+ \Lambda_K K^{\dagger} K\vert_D.
\end{equation}
A non-zero vev of the scalar component of $K$ would change the
normalization of the $\Phi$ kinetic term, thus we set it to zero.
However a term like (\ref{KK}) will be useful to us
in what follows. Note that
$K$ carries arbitrary
$U(1)$ charges. If $K$ were to appear in the
superpotential  the number of powers of $K$ would be dictated by
its $U(1)$ charges. However, the theory must
be invariant under the arbitrary
assignment of
charge, and the only function of $K$ which has this
property is the constant function. Hence there is no
$K$ dependence in the superpotential.
The $D$-terms involving $K$ will of course in general be
renormalized by an arbitrary,
non-holomorphic  function of the fields in the model (wavefunction
renormalization).

It is important to note that the relationship
between the superpotential and the potential of any model
depends on the coefficients of the $D$-terms. This is because the
auxiliary fields $F$ must be eliminated in order to obtain
the potential as a function of propagating fields only.
Since the coefficient functions of the $D$-terms are not
holomorphic in the fields, this means that an exact result for the
superpotential does not necessarily yield an exact result for the
potential.

\subsection{ N flavor WZ model}

The arguments of section 2.1 may be extended to an
N flavor WZ model with Lagrangian

\begin{equation}
\label{LWZN}
L_0 ~=~ \sum_{i}^N (K^{\dagger}_iK _i+ 1)
\Phi^{\dagger}_i \Phi _i \vert_D ~+~ \lambda_{i}
\Phi_i \vert_F + m_{ij}
\Phi_i \Phi_j \vert_F + g_{ijk}
\Phi_i \Phi_j \Phi_k \vert_F ,
\end{equation}
where $i,j,k = 1, ..,N$.
We may again embed the model in a larger
Lagrangian in which the couplings are themselves
chiral superfields with infinite $D$-terms.
The theory now has $N$ global $U(1)$ symmetries
in addition to $U(1)_R$. The fields transform as

\begin{equation} \begin{array}{cccc}
 & U(1)_k &  {}  & U(1)_R\\
\Phi_i &  \delta_{ik} & & 1\\
\lambda_{i} & -\delta_{ik} & & 1 \\
m_{ij} &  -(\delta_{ik} + \delta_{jk}) & & 0\\
g_{ijl} &  -(\delta_{ik} + \delta_{jk} + \delta_{lk}) & & -1\end{array}
\end{equation}

\noindent  The effective superpotential is now constrained to the form

\begin{equation} W_{eff} ~=~ m_{ij} \Phi_i \Phi_j \hspace{.3cm}
f\left( \frac{\lambda_{a} \Phi_a}
{m_{ij} \Phi_i \Phi_j}  ; \frac{g_{abc}
\Phi_a \Phi_b\Phi_c} {m_{ij} \Phi_i \Phi_j} \right)  .
\end{equation}

\noindent  The superpotential is independent of $K$ since $K$
has arbitrary $U(1)_R$ charge.
Using similar arguments as for the one
flavor model we see that the superpotential is
again not renormalized and is exact even at the
non-perturbative level.

\subsection{Soft SUSY breaking terms in the WZ model}

Now we will generalize the previous analysis to allow for
soft breaking of supersymmetry.
Consider the following model which is a
subclass of the $N$ flavor WZ model discussed in section 2.2

\begin{eqnarray}
\label{LOR}
 L & ~=~ &  (K_0^{\dagger}K_0 + 1)
  \Phi^{\dagger}  \Phi \vert_D
+ m_0 \Phi ^2 \vert_F + g_o \Phi^3 \vert_F \nonumber \\
&& ~+~ \Lambda_m( m_i^{\dagger}m_i \vert_D + \alpha_m m_0 \vert_F
+ \beta_m m_1 m_2 \vert_F + \gamma_m m_{1}^2 m_0 \vert_F )  \nonumber \\
&& ~+~ \Lambda_g( g_i^{\dagger}g_i \vert_D + \alpha_g g_0 \vert_F
+ \beta_g g_1 g_2 \vert_F + \gamma_g g_{1}^2 g_0 \vert_F)  \nonumber \\
&& ~+~ \Lambda_K( K_i^{\dagger}K_i \vert_D + \alpha_K K_0 \vert_F
+ \beta_K K_1 K_2 \vert_F + \gamma_K K_{1}^2 K_0 \vert_F) \nonumber \\
&& ~+~ h.c.
\end{eqnarray}
where  $m_i, g_i, K_i$ are
superfields with $i= 0,1,2$ and
the couplings $\alpha$, $\beta$ and $\gamma$ are complex.
Since this model is a specific example  of the
$N$ flavor WZ model we know that its superpotential
 is not renormalized. The potentials for
$m_i$, $g_i$ and $K_i$ are simply those of the
O'Raifeartaigh model \cite{OR} \footnote{We can actually
eliminate all of the $c_i$, $i = 1,2$ fields from (\ref{LOR})
and still have spontaneous SUSY breaking.}
and can be minimized by
\begin{eqnarray}
\langle F_{c_0} \rangle   ~&=&~ \alpha_c^{\ast}  \nonumber \\
\langle A_{c_0} \rangle ~&\neq&~ 0 \nonumber \\
\langle A_{c_i} \rangle ~&=&~ 0 ~~~~~~ i = 1,2 ~~~   ,
\end{eqnarray}
where the subscript $c_i \equiv m_i, g_i, K_i$.
The $\langle  F \rangle$ vevs  spontaneously break SUSY.
The potential is flat at ${\cal O}(\Lambda_c)$ in
$A_{m_0}$, $A_{g_0}$ and $A_{K_0}$ so we may
choose these vevs arbitrarily when we take the
limit $\Lambda_c \rightarrow \infty$.
We shall choose these vevs equal to
the desired one flavor WZ couplings
and also take $\langle A_{K_0}\rangle = 0$.

After taking the limit of all $\Lambda_c$ to infinity,
we obtain the following $F$-terms in the effective theory
\begin{equation}
\label{Lembed}
\left[ \langle A_{m_0}\rangle \Phi^2 \vert_F ~+~
\langle A_{g_0}\rangle\Phi^3\vert_F ~+~ h.c. \right]
{}~+~ 2Re( \langle F_{m_0} \rangle A^2 ~+~ \langle F_{g_0} \rangle A^3 )
\end{equation}
where $A$ is the scalar component of the chiral field $\Phi$.

Again, since the Langrangian we started with in (\ref{LOR})
was exactly supersymmetric, and of the form (\ref{LWZN}),
the terms in (\ref{Lembed}) are not renormalized --
they are {\it exact}.   While (\ref{Lembed}) appears to violate
SUSY explicitly, albeit softly, we actually obtained it
as the limit of a model in which SUSY is only
spontaneously broken. Hence the persistence of the exact result.

The $D$-term in the bare Lagrangian now
generates a soft SUSY breaking term

\begin{equation}
\langle F^{\ast}_{K_0} F_{K_0} \rangle~ A^{\ast} A  ~  .
\end{equation}

\noindent While we cannot specify the exact renormalized $D$-term,
it is highly plausible that if
\nobreak $\langle F^{\ast}_{K_0} F_{K_0} \rangle$
is large enough this breaking term
will persist in the effective theory.

We have induced the following soft
SUSY breaking terms (writing $A = a + ib$):
\begin{eqnarray}
\label{soft}
&{}&  \langle F_{K_0}^* F_{K_0} \rangle ~(a^2 + b^2) \nonumber \\
&{}& Re \langle F_{M_0} \rangle ~(a^2-b^2)~ \nonumber \\
&{}& Im \langle F_{M_0} \rangle ~ 2ab~ \nonumber \\
&{}& Re \langle F_{g_0} \rangle ~ (a^3-3ab^2)~ \nonumber \\
&{}& Im \langle F_{g_0} \rangle ~(b^3-3a^2b) ~  .
\end{eqnarray}
We note that these exhaust the types of soft SUSY breaking
terms which are consistent with the perturbative
non-renormalization result \cite{GG}
\footnote{In \cite{GG} the imaginary terms were not obtained because  only
real couplings were considered.}. Each term in (\ref{soft}) is exact, with the
exception of the first, which arises from a $D$-term.

Allowing vevs of the $F$ components of the coupling constants
can lead to additional $D$-terms which are holomorphic in $\phi$
and hence mimic terms in the softly broken WZ model.
Specifically, terms of the form
\begin{equation}
\label{Dstuff}
  f ( g^\dagger g )  (~ g \phi m^\dagger ~)^l \vert_D
\end{equation}
with $l \geq 1$ will appear in the effective Lagrangian. These terms
are consistent with the $U(1)$ symmetries and lead to $F$-terms
in $\phi$ when $\langle F_{g,m} \rangle ~\neq~ 0$.
To arrive at the form (\ref{Dstuff}), we have eliminated terms
which are singular as $m$ or $\phi~ \rightarrow ~0$, or which
can become singular at weak coupling (e.g. $e^{- m / g \phi}$).
The terms remaining in (\ref{Dstuff}) can be seen to
receive contributions
within perturbation theory from specific supergraphs,
and may also receive non-perturbative contributions
of the form $e^{- 1 /( g^\dagger g )}$. Thus we cannot
exactly determine the functions $ f ( g^\dagger g )$.
However, simple power counting tells us that
the term linear in $\phi$ has at worst logarithmically
divergent contributions and that higher order terms
have finite contributions. Therefore these additional
terms do not spoil the non-renormalization results
of \cite{GG}.

\section{SUSY QCD}

Soft SUSY breaking terms may also be
introduced into SQCD. The SQCD Lagrangian is

\begin{eqnarray}
\label{LYM}
L & ~=~ &  \frac{1}{4} ~
( \tau W^{a \alpha}W_{a \alpha}
\mid_F  ~+~
\tau^{\dagger} \bar{W}_{a \dot{\alpha}} \bar{W}^{a \dot{\alpha} })
\mid_F  ) \nonumber \\
&&  ~+~ (1+K^{\dagger} K) Q^{\dagger}e^VQ \mid_D
{}~+~ (1 +\tilde{K}^{\dagger} \tilde{K})
\tilde{Q}^{\dagger} e^{-V} \tilde{Q}  \mid_D \nonumber \\
&& ~+~ m \tilde{Q}Q \mid_F  ~+~
m^{\dagger} Q^{\dagger} \tilde{Q}^{\dagger} \mid_F    .
\end{eqnarray}
Following our procedure in the WZ model we promote
$\tau$, $m$, $K$ and $\tilde{K}$ to the status of chiral superfields
and then couple each field to an O'Raifeartaigh model
allowing the $F$ and scalar components of each field
to acquire a non-zero vev. After fixing the
coupling constant fields we obtain the SQCD
Lagrangian plus the  SUSY breaking terms

\newpage

\begin{eqnarray}
\label{dLYM}
\Delta L~ = ~ \langle F_{K}^{\ast}F_{K}\rangle \mid A_Q\mid^2 ~ + ~
 \langle  \tilde{F}_{\tilde{K}}^{\ast}  \tilde{F}_{\tilde{K}} \rangle
\mid \tilde{A}_{\tilde{Q}} \mid^2 ~  + ~
2Re(\langle F_{m}\rangle \tilde{A}_{\tilde{Q}} A_Q) \nonumber  \\
 ~ +  ~  \frac{1}{4} \langle F_{\tau} \rangle
\lambda^{\alpha} \lambda_{\alpha} ~  + ~
\frac{1}{4}  \langle F_{\tau}^{\ast}\rangle \bar{\lambda}_{\dot{\alpha}}
\bar{\lambda}^{\dot{\alpha}}  .
\end{eqnarray}
Writing the squark fields as $(a_Q ~+~  i b_Q )$
we explicitly induce the soft SUSY breaking parameters
\begin{equation}
\label{SSB}
\begin{array}{c}
\langle F_{K}^{\ast} F_K\rangle ~(a_Q^2 + b_Q^2) \\ \\
\langle \tilde{F}_{\tilde{K}}^{\ast} \tilde{F}_{\tilde{K}} \rangle
(\tilde{a}_{\tilde{Q}}^2 + \tilde{b}_{\tilde{Q}}^2) \\ \\
\langle F_{\tau}\rangle  \lambda^{\alpha} \lambda_{\alpha}  ~+~
\langle F_{\tau}^{\ast}\rangle \bar{\lambda}_{\dot{\alpha}}
\bar{\lambda}^{\dot{\alpha}}
\end{array}
\end{equation}

\noindent which have the effect of giving masses
to the gaugino and squark fields.

The model possesses an anomaly free global
$SU(N_f)_L \otimes SU(N_f)_R \otimes U(1)_V \otimes U(1)_R$
symmetry where the fields transform under  the $U(1)_R$ group as

\begin{equation} \begin{array}{ccc}
& & U(1)_R\\
W & & 1 \\
\tau & & 0\\
 Q &  & (N_f - N_c)/ N_f \\
\tilde{Q} & &  (N_f - N_c)/ N_f \\
m & & 2N_c/N_f \\
K & & {\rm arbitrary} \end{array} \end{equation}

As a simple example, consider the case of $N_c > N_f$ .
The effective superpotential
(or more precisely, the $F$-term part of $L_{eff}$)
is therefore determined by the symmetries to be of the form

\begin{equation}
\label{WSQ}
W_{eff} = f\left( \tau, \frac{det(\tilde{Q}
Q)^{1/(N_f-N_c)}}{W^{\alpha}W_{\alpha} },
m^{N_f(N_c-N_f)/N_C} {\rm det}(\tilde{Q} Q) \right) W^{\beta} W_{\beta}  ~+~
h.c.\label{fred}
\end{equation}
(If $N_f \geq N_c$ then objects other than those in
(\ref{WSQ}) can appear \cite{S2}.)
Again the superpotential is not dependent on $K$ since it has arbitrary charge.
The contribution to the potential at lowest order in a derivative
expansion is the term reported in \cite{ADS}:
\begin{equation}
\label{WSQ1}
\frac{\Lambda^{(3N_c-N_f)/(N_c-N_f)}}
{{\rm det}(\tilde{Q} Q)^{1/(N_c-N_f)}}  .
\end{equation}
The dimensional factor $\Lambda$ must be proportional to
$\Lambda_{SQCD}$, which is the only scale in the problem other than $m$,
which has $R$ charge.
$\Lambda_{SQCD}$ can be related to the gauge coupling constant:
$\Lambda_{SQCD} \sim \mu~ exp ( - \langle  A_\tau \rangle / ~b )$,
where $b$ is determined
by the $\beta$-function. The general function of
$\tau$ which appears in (\ref{WSQ1}) is then
determined to be $exp ( - \tau  / b )$,
and can be straightforwardly evaluated when $\langle  F_{\tau} \rangle \neq 0$.
The functional
form of the superpotential is unchanged when we allow the $F$-terms of $M$,
$K$ and $\tau$ to have non-zero vevs and
therefore we expect many of Seiberg's results in
SQCD to  apply in theories with soft SUSY breaking terms.

As in the case with the WZ model, additional terms which are
holomorphic in the physical (non-coupling constant) fields
can arise from soft breaking. In particular, the following
$D$-terms are allowed by the symmetries
(we consider the massless case $m = 0$):
\begin{equation}
\label{Dstuff1}
f~( \tau, \tau^\dagger,  ~det(\tilde{Q}
Q)^{1/(N_c-N_f)} ~W^{\alpha} W_{\alpha}~  )~ \vert_D   .
\end{equation}
When $\langle f_\tau \rangle \neq 0$, this leads to additional
superpotential terms.
Terms of the form (\ref{Dstuff1})
are not generated within perturbation theory but may
receive nonperturbative contributions.
They are qualitatively similar to terms already
present in (\ref{WSQ}) which couple the quark and gauge superfields.

We again note that, due to the unknown renormalization of the $D$-term,
the potential is not exactly determined, even when the
superpotential is. In the limit where the theory is weakly coupled, the
renormalization of the $D$-terms can be computed within perturbation
theory, and hence the potential determined (albeit not exactly)
from the superpotential. However, if the model is strongly coupled
the coefficient functions of the $D$-terms could deviate considerably
from their classical values and hence drastically
alter the relation between the potential
and the superpotential. Thus it does not appear that any
{\it exact} results can be obtained regarding the {\it potential},
despite the dramatic results concerning the superpotential.

Modulo the caveats of the preceding paragraph, it is expected \cite{ADS}
that SQCD with $m = 0$ has no vacuum for $N_c > N_f$. As a simple example
of the effect of soft-breaking, consider allowing
\begin{equation}
\langle F_{K}^{\ast} F_K\rangle ~=~
\langle \tilde{F}_{\tilde{K}}^{\ast} \tilde{F}_{\tilde{K}} \rangle
{}~\neq~ 0,
\end{equation}
with no other soft-breaking in the model. The mass terms
generated for the
squark fields (see (\ref{SSB}))
then prevent the runaway behavior of the exactly supersymmetric
model, and allow it to have a vacuum. Allowing gaugino masses
through $\langle F_{\tau} \rangle \neq 0$ leads to the following
additional term in the potential
\begin{equation}
-
\frac{ \langle F_{\tau}  \rangle }{b} ~
\frac{\Lambda^{(3N_c-N_f)/(N_c-N_f)}}
{{\rm det}(\tilde{A}_{\tilde{Q}} A_Q)^{1/(N_c-N_f)}}
\end{equation}
which does not by itself alter the runaway behavior.

\section{Discussion}

We have demonstrated that the exact results of \cite{S1} which rely
on holomorphicity properties of the Wilsonian superpotential
can be generalized to models in which supersymmetry is broken
softly. This yields a large new class of exact results for
non-supersymmetric models.

As a particularly interesting application of our results,
vacuum expectation values of $F_{K}$ and
$F_{\tau}$ can be used to give masses
to the scalar components of matter chiral superfields (e.g. squarks,
selectrons) and gauginos (gluinos, photino) respectively.
This means that we can recover QCD and QED as softly broken
versions of their supersymmetric counterparts SQCD and SQED,
with exact results intact. Unfortunately, many of the exact results are
useful primarily to determine the vacuum structure of the scalar sector
of the model, which is precisely what is decoupled in the limit
that $\langle F_{K} \rangle$ is taken to infinity. However,
the coefficient of $W^\alpha W_\alpha \vert_F$, which is related to the
$\beta$-function, is also holomorphic
and does not decouple.

There are supersymmetric results
used in Seiberg's analysis in
\cite{S2} that are not left intact
by  the spontaneous breaking of SUSY.
In particular, correlators of lowest components of chiral
superfields
\begin{equation}
\label{corr}
G( x_1, ~... ~, x_n) ~=~ \langle  0 \vert~ T \left( A(x_1) ~... ~A(x_n)
\right) \vert 0 \rangle
\end{equation}
are no longer necessarily position independent
or holomorphic. These properties rely on the assumption
that the vacuum is supersymmetric, and hence
annihilated by the SUSY symmetry generators
$ Q_\alpha, \bar{Q}_{\dot{\alpha}}  $ .
This clearly no longer holds in a theory in which
SUSY is spontaneously broken.
The modification of the supersymmetric Ward-Takahashi
identities used to derive the properties of (\ref{corr})
(see \cite{CERN}, and references therein)
is likely to lead to
modifications of some of the results in \cite{S2}.
We leave this and other issues involving applications
of our results to future investigation.

\vskip 1.0 in
\centerline{\bf Acknowledgements}
\vskip 0.1in

The authors would like to thank
P. Hernandez,  G. Moore, S. Selipsky and R. Sundrum
for useful comments or discussions, and N. Seiberg
for some important comments about $D$-terms.
This work was supported under DOE contract DE-AC02-ERU3075.

%

\def\ap#1#2#3{           {\it Ann. Phys. (NY) }{\bf #1}, #2 (19#3)}
\def\apj#1#2#3{          {\it Astrophys. J. }{\bf #1}, #2 (19#3)}
\def\apjl#1#2#3{         {\it Astrophys. J. Lett. }{\bf #1}, #2 (19#3)}
\def\app#1#2#3{          {\it Acta Phys. Polon. }{\bf #1}, #2 (19#3)}
\def\ar#1#2#3{     {\it Ann. Rev. Nucl. and Part. Sci. }{\bf #1}, #2  (19#3)}
\def\com#1#2#3{          {\it Comm. Math. Phys. }{\bf #1}, #2 (19#3)}
\def\ib#1#2#3{           {\it ibid. }{\bf #1}, #2 (19#3)}
\def\nat#1#2#3{          {\it Nature (London) }{\bf #1}, #2 (19#3)}
\def\nc#1#2#3{           {\it Nuovo Cim.  }{\bf #1}, #2 (19#3)}
\def\np#1#2#3{           {\it Nucl. Phys. }{\bf #1}, #2 (19#3)}
\def\pl#1#2#3{           {\it Phys. Lett. }{\bf #1}, #2 (19#3)}
\def\pr#1#2#3{           {\it Phys. Rev. }{\bf #1}, #2 (19#3)}
\def\prep#1#2#3{         {\it Phys. Rep. }{\bf #1}, #2 (19#3)}
\def\prl#1#2#3{          {\it Phys. Rev. Lett. }{\bf #1}, #2 (19#3)}
\def\pro#1#2#3{          {\it Prog. Theor. Phys. }{\bf #1}, #2
(19#3)}
\def\rmp#1#2#3{          {\it Rev. Mod. Phys. }{\bf #1}, #2 (19#3)}
\def\sp#1#2#3{           {\it Sov. Phys.-Usp. }{\bf #1}, #2 (19#3)}
\def\sjn#1#2#3{           {\it Sov. J. Nucl. Phys. }{#1}, #2 (19#3)}
\def\srv#1#2#3{           {\it Surv. High Energy Phys. }{#1}, #2
(19#3)}
\def\tp{these proceedings}
\def\zp#1#2#3{           {\it Zeit. fur Physik }{\bf #1}, #2 (19#3)}
%

\end {document}